# Biocompatible Two-dimensional Titanium Nanosheets for Efficient Plasmonic Photothermal Cancer Therapy

*Zhongjian Xie[a,1], Shiyou Chen[a,1], Quan Liu[c,d], Zhitao Lin[a], Jinlai Zhao[d], Taojian Fan[a], Dou Wang[c,d], Liping Liu[b,\*], Shiyun Bao[b], Dianyuan Fan[a], and Han Zhang[a,\*]*

[a]Shenzhen Engineering Laboratory of Phosphorene and Optoelectronics, Collaborative Innovation Center for Optoelectronic Science & Technology, International Collaborative Laboratory of 2D Materials for Optoelectronics Science and Technology of Ministry of Education, Shenzhen University, Shenzhen 518060, China

[b]Department of Hepatobiliary and Pancreatic Surgery, Shenzhen People's Hospital, Second Clinical Medical College of Jinan University, Shenzhen, Guangdong Province, P. R. China.

[c]Integrated Chinese and Western Medicine Postdoctoral Research Station, Jinan University, Guangzhou 510632, China

[d]College of Materials Science and Engineering, Shenzhen University, Shenzhen 518060, PR China

[\*] To whom correspondence should be addressed.
E-mail: hzhang@szu.edu.cn

[1] Z.X. and S.C. contributed equally to this work.

**Abstract**

Photothermal therapy (PTT) based on two-dimensional (2D) nanomaterials has shown significant potential in cancer treatment. However, developing 2D nanomaterial-based photothermal agents with good biocompatibility and high photothermal conversion efficiency (PTCE) remains a key challenge. Titanium and its alloys have been widely employed as biomedical materials based on their biocompatibility. In this work, the elemental Ti based plasmonic photothermal therapy (Ti-PPTT) is demonstrated. Using the liquid-phase exfoliation (LPE), it was shown that metallic Ti can be fabricated into the 2D nanosheets (NSs), similar to exfoliating other layered 2D materials. Furthermore, the 2D Ti NSs exhibited good biocompatibility, high extinction coefficient of 20.8 $Lg^{-1}cm^{-1}$ and high PTCE of 73.4%, owing to localized surface plasmon resonances (LSPR); which is significantly higher than other photothermal agents, including Au (21%), $MoS_2$ (24.4%), BP (28.4%) and $Ti_3C_2$ MXene (30.6%). Interestingly, the Ti NSs can be degraded into $TiO_2$ in water-based solution and made themselves hydrophilic in the meantime. Consequently, 2D Ti-PPTT exhibited a notable therapeutic effect in a human hepatocellular carcinoma model without side effects. Our study could pave a new avenue for PTT using metal Ti and arouse a wide interest in the potential efficient PTT for other elemental transition metals owing to their LSPR.

# 1. Introduction

Materials used for biomedical applications cover a very broad range and must exhibit specific properties depending on application. The prerequisite of all biomedical materials however, is biocompatibility. Titanium (Ti) and its alloys are among the newest and most attractive metallic biomaterials because of their biocompatibility.[1,2] In both medical and dental fields, Ti and its alloys have been demonstrated as a successful class of biomedical materials, which do not give rise to adverse effects locally or systemically.[3] In medicine, they are frequently used as implant devices to replace failed hard tissues, including artificial knee joints, artificial hip joints, bone plates, cardiac valve prostheses and artificial hearts.[4] For dentistry devices, Ti and its alloys are used for bridges, crowns, overdentures and dental implant prosthesis components.[5]

Pure metallic Ti is considered to be one of the best biocompatible metallic materials. The main physical properties of Ti responsible for its biocompatibility are: high corrosion resistance,[3,4] low electronic conductivity, its thermodynamic state at physiological pH values, an isoelectric point of the oxide of 5–6, and low ion-formation tendency in aqueous environments.[2] Moreover, the inert oxide layer covered surface is only slightly negatively charged at physiological pH. The dielectric constant of Ti, is comparable to that of water resulting in the coulomb interaction of the charged species being similar to that in water.

Photothermal therapy (PTT) based on the stimulus of near-infrared (NIR) laser irradiation is a noninvasive therapeutic modality compared with traditional therapies for cancer treatment.[6] In the past decade, many photothermal agents, such as noble metal nanomaterials (NMs),[7] carbon-based NMs,[8] copper sulfide,[9] rare earth compounds,[10] and many organic nanoparticles and polymers,[11] have been widely explored. Recently, 2D NMs, including palladium (Pd),[12] $MoS_2$,[13–16] $WS_2$,[17,18] and black phosphorus (BP),[19] have attracted attention as photothermal agents due to their successful photothermal performances. An efficient photothermal agent should not only possess an enhanced absorption in the NIR region, but also a high photothermal conversion efficiency (PTCE). Importantly, the biocompatibility prerequisite for photothermal agents should also be readily satisfied.

Plasmonic NMs with localized surface plasmon resonances (LSPR) in the NIR region have been extensively explored as photo-sensitizers for PTT over the past decade.[20–22] Compared with conventional photo-sensitizers, plasmonic nanoparticles not only possess stronger NIR light absorption, owing to LSPR, but also exhibit higher photostability.[22] Plasmonic NMs are composed of various materials, such as metals (Au, Ag), and semiconductors. In view of the high biocompatibility of Ti alloys, they have not only been widely investigated for biocompatible implants, but also for emerging PTT-based applications. Up to now, several titanium compounds, such as titanium carbide,[23–26] titanium oxide,[27–29] titanium nitride,[20,30,31] and $Ti_xTa_{1-x}S_yO_z$,[32] have been investigated as photothermal agents for cancer therapy. They not only possess excellent biocompatibility, but also exhibit high photothermal performance owing to their plasmon resonance.

Titanium carbide ($Ti_3C_2$) is an example of an MXene, which is a new class of 2D early transition-metal carbides. $Ti_3C_2$ MXene exhibited high antibacterial efficiency with growth inhibition of 97.7%.[33] The $TiO_2$-$Ti_3C_2$ nanocomposite has also been shown to be a promising biosensor for the detection of $H_2O_2$ with high sensitivity.[34] $Ti_3C_2$ based field-effect transistors, with high biocompatibility, were used to detect dopamine, and to monitor spiking activity in hippocampal neurons.[35] $Ti_3C_2$ quantum dots (QDs) have been shown to have great potential for multicolor cellular imaging, and in the optical field in general.[36] These investigations revealed that $Ti_3C_2$ has great potential in environmental and biomedical applications

Notably, $Ti_3C_2$ MXene was reported as an effective 2D light-to-heat conversion material with high photothermal efficiency under sunlight irradiation.[23] Moreover, a novel photothermal agent based on ultrathin $Ti_3C_2$ MXene showed strong optical absorption in the NIR region.[24] $Ti_3C_2$ MXene also demonstrated highly effective photothermal ablation of tumors.[25] In spite of the high PTT performance of $Ti_3C_2$ MXene, its fabrication is a hazardous and time-consuming process, using HF pretreatment. Fortunately, a new and simple fluorine-free fabrication method has been proposed and photoacoustic (PA) imaging and notable tumor ablation have been demonstrated simultaneously.[26] The strong absorption and high conversion efficiency of $Ti_3C_2$ MXene originate from the LSPR effect of its semimetal character. All of the highlighted reports support the potential of $Ti_3C_2$ based MXene application in biomedicine, especially for photothermal cancer treatment.

In the LSPR enhanced photothermal effect, heat generation is particularly high in the case of metal nanoparticles, given that LSPR is a collective motion of numerous electrons.[37] In comparison, heat generation is much weaker for semiconductor nanoparticles, since heat dissipation occurs through the creation of single mobile electrons and holes. Consequently, metallic NMs with plasmon resonance can be developed as effective and promising photothermal agents.

In this work, based on the widely used biocompatible material Ti, we report a photothermal agent of 2D metal Ti nanosheets (NSs), which exhibits excellent biocompatibility and highly efficient photothermal performance both *in vitro* and *in vivo* (**Figure 1**). It is interesting to find that bulk Ti can be processed to give the 2D form (Ti NSs) using liquid-phase exfoliation (LPE), as has been demonstrated for other layered 2D materials such as black phosphorus. The Ti NSs showed high photothermal performance, evaluated by both extinction coefficient and PTCE. Surface modification with PEG also led to colloidal stability in water, saline and physiological conditions. *In vitro* and *in vivo* experiments were conducted to systematically investigate the toxicity of the Ti NSs, and a hepatocellular carcinoma model was established to evaluate the potential of Ti NSs for photothermal tumor therapy.

## 2. Results and Discussion

## Fabrication

Non-layered 2D materials, the other class of 2D materials, have been of great interest in recent years because of their useful properties.[38,39] Distinct from 2D layered materials, non-layered materials have chemical covalent bonds in three dimensions. After exfoliation, the surfaces of non-layered 2D materials present dangling bonds, which render their surfaces highly chemically active and enhance their catalytic capability, sensing, and carrier transfer. These features are not present in the case of 2D layered materials. Given these advantageous properties, many well-established techniques have been developed to synthesize non-layered 2D materials, such as template directed synthesis (TDS),[40,41] the lamellar hybrid intermediate method,[42,43] and van der Waals epitaxial (vdWE) growth.[44,45]

LPE can produce 2D structures through the choice of an appropriate solvent, with the aid of sonication. The liquid phase can be considered the most efficient and versatile as it allows for preparation of many materials,[46] and is particularly suitable for biomedical purposes.[26] LPE can also be used to produce specific surface functionalization, which is particularly advantageous for stabilization of the 2D nanosheets in solution.[46] However, using LPE to exfoliate non-layered 2D materials is rarely investigated.

2D layered materials can be exfoliated through LPE because of their clear anisotropic bonding nature in three dimensions; strong covalent bonding interactions intralayer (the in-plane direction) and weak physical van der Waals' forces between layers (the out-of-plane direction). While, for non-layered 2D materials, the chemical covalent bonds in three dimensions still show some anisotropic bonding character. Based on this anisotropy, they can also be exfoliated into a 2D form using the LPE method. For instance, the large-area freestanding single layers of non-layered ZnSe with four-atom thickness were exfoliated using LPE.[43] And ultrathin $CoSe_2$ nanosheets exfoliated using LPE can be used to efficiently catalyze the oxygen evolution reaction.[47] In our previous work, the non-layered VI A elements selenium (Se) and tellurium (Te) were exfoliated into 2D form using LPE (not published).

It is clear that Ti metal is a non-layered material. It has a *hcp* crystal structure, which presents mechanical anisotropy. Therefore, it can theoretically be exfoliated using LPE, and interestingly, our attempts have found that typical 2D Ti NSs can be obtained (**Figure 2**) using the typical LPE method (**Figure 1**). As in exfoliation of other 2D layered materials, isopropyl alcohol (IPA) was chosen as the solvent for this exfoliation. Compared with other liquid choices such as N-methyl-2-pyrrolidone (NMP),[48] this solvent can be easily evaporated to ensure the surface of the obtained Ti NSs is free of impurities for subsequent biological experiments.

## Characterization

To obtain 2D Ti NSs with the desired dimensions, dispersions of Ti NSs subjected to varying degrees of centrifugal force were characterized using transmission electron microscopy (TEM) and atomic force microscopy (AFM). **Figure 2a** shows the dimensions of the Ti NSs obtained from the supernatant of a dispersion subjected to a centrifugal force of 2000 ×g. The TEM image shows a typical Ti NS lateral size of less than 50 nm. Additionally, some Ti QDs can also be observed. The AFM image (**Figure 2b**) shows that the thickness of the Ti NSs falls within the range of 1–5 nm. Clear lattice fringes with inter-atomic d-spacing of 0.21 nm, 0.24 nm and 0.26 nm can be observed (**Figure 2c**). The selected-area electron diffraction (SAED) pattern suggests the typical crystalline features of α-Ti.[49] Fast Fourier Transformation (FFT) of the HRTEM image shows the expected crystallographic lattice reflections of the Ti NSs.

The chemical composition and crystal phase of bulk Ti and Ti NSs were characterized using X-ray photoelectron spectroscopy (XPS) and X-ray diffraction (XRD). In **Figure 2d**, the two XPS peaks at 458.58 eV and 454.33 eV are ascribed to the Ti $2p_{3/2}$ and Ti $2p_{1/2}$ orbitals, respectively, for both bulk Ti and Ti NSs.[50,51] The XPS peak at 464.33 eV indicates the presence of $TiO_2$.[52] The XRD patterns for both the bulk Ti and exfoliated Ti NSs exhibited typical Ti crystalline diffraction (**Figure 2e**).[51,53] In **Figure 2f**, the bulk Ti and Ti NSs exhibited similar Raman peaks. A sharp peak at 140.5 $cm^{-1}$ was observed for bulk Ti, which is assigned to the zone-center $E_{2g}$ mode, which can be directly assigned to *hcp* titanium.[54] For exfoliated Ti NSs, a peak at 143.8 $cm^{-1}$ indicates a Raman shift for the ultrathin Ti NSs. The peaks at ~200 $cm^{-1}$ for both bulk Ti and exfoliated Ti NSs arise as a result of $TiO_2$.[55]

All characterizations confirm the dimension, composition and crystal features of the exfoliated Ti NSs. It can therefore be concluded that the LPE method can be used to exfoliate non-layered metal materials with anisotropic crystal structures in three dimensions; and is not limited to layered materials showing more significant contrast between dimensions, for example, between covalent bonding and van der Waals' interactions. This widely expands the scope of the LPE method and means more 2D materials with unknown properties can be explored. The exfoliated Ti NSs with relatively small size (<50 nm lateral size and ~3 nm in thickness) can be used in PTT.

The stability of the Ti NSs in physiological medium was enhanced by surface coating with polyethylene glycol (PEG). The success of this coating was validated by Fourier transform infrared spectroscopy (FTIR) and scanning transmission electron microscopy (STEM) with energy dispersive X-ray spectroscopy (EDS) mapping (**Figure 3**). For the naked Ti NSs, the intense band at ~3400 $cm^{-1}$ was reported to be resulted from the OH stretching. The composition of $TiO_2$ in the surface of Ti NSs has been confirmed in the former XPS and Raman measurements (**Figure 2**). Thus, the OH group was produced by $TiO_2$ and made the surface of Ti NSs hydrophilic.[56,57] Compared with the spectrum of naked Ti NSs, two additional FTIR peaks emerged at ~1100 $cm^{-1}$ and ~2900 $cm^{-1}$ for the PEGylated Ti NSs. Comparison with the pure PEG spectrum (**Figure 3a**) shows that these two peaks can be assigned PEG. The absorption band at ~2900 $cm^{-1}$ is attributed to the C-H vibration, and that at ~1100

cm$^{-1}$ is assigned to C-O stretching in the PEG unit, indicating the successful PEGylation of Ti NSs. The thickness of PEGylated Ti NSs from the AFM image is ~10 nm (**Figure 3b**), which is slightly thicker compared with the naked Ti NSs (**Figure 3b**). Dynamic Light Scattering (DLS) measurements showed that the size distribution of PEGylated Ti NSs shifted to larger sizes compared with naked Ti NSs (**Figure 3c**). Moreover, the STEM mapping shows the co-localization of four different elements: Ti, C, O and N. The C, O and N contributions come from the surface coating of PEG.

## Photothermal performance

Extinction coefficient defines how strongly materials absorb light. A high extinction coefficient is a prerequisite for effective photothermal agents. To characterize this value for Ti NSs, their optical absorption at different concentrations was measured. The photograph of Ti NSs dispersed in water at concentrations of 10, 25, 50 and 100 ppm is shown in **Figure 4a**. The concentration (*C*) was determined using inductively coupled plasma atomic emission spectroscopy (ICP-AES). The 100 ppm dispersion was completely opaque due to the strong absorption of Ti NSs. The optical absorption spectra of Ti NSs at different concentrations are shown in **Figure 4b**. They show a broad and strong absorption band spanning from the UV to NIR regions, similar to those of other non-metallic, layered 2D NMs, such as GO[8], MoS$_2$,[13] WS$_2$[58] and BP.[19] Strong absorption in the NIR region is necessary to take advantage of the NIR transparent window (750–1000 nm) of biological tissue for PTT.[59] The normalized absorbance over the light path length of the measurement cuvette (A/L, where A is absorbance and L is length of the light path) at 808 nm for different concentrations, was determined (**Figure 4d**). The extinction coefficient (k) at 808 nm was then calculated to be 20.8 Lg$^{-1}$cm$^{-1}$, according to the Lambert-Beer law (A/L=k*C*). This is 5.3 times greater than that of AuNRs (3.9 Lg$^{-1}$cm$^{-1}$).[19] Compared with previously reported 2D NMs, the extinction coefficient of the Ti NSs is higher than that of GO NSs (3.6

Lg$^{-1}$cm$^{-1}$),[8] and even higher than that of popular photothermal agent BPQDs (14.8 Lg$^{-1}$cm$^{-1}$).[19] Although it is slightly lower than the extinction coefficient of Ti$_3$C$_2$ (25.2 Lg$^{-1}$cm$^{-1}$),[25] it still shows significant potential in photothermal conversion efficiency (PTCE), another important property for evaluating photothermal agents.

Different concentrations of Ti NSs in water were exposed to an 808 nm NIR laser with power density of 1.0 W.cm$^{-2}$, which obeys the maximum permissible exposure (MPE) for skin of 1 W.cm$^{-2}$ (American National Standard for Safe Use of Lasers, ANSI Z136.1−2007).[60,61] The temperature of the Ti NSs dispersion was measured as irradiation time increased (**Figure 4c**). The power density of the laser was calibrated before irradiation. At a low concentration of 50 ppm, the temperature increased from 25 ℃ to 58.5 ℃ after 600 s of irradiation. The PTCE ($\eta$) is the key parameter for evaluating photothermal performance. PTCE of the Ti NSs was determined to be up to 73.4 % (**Figure 4e**) using a standard method,[62] indicating that the Ti NSs can efficiently convert NIR light into heat. The determined value was significantly higher than other photothermal agents, such as Au nanoparticles (21 %),[63,64] MoS$_2$ (24.4 %),[14] BPQDs (28.4 %),[19] recently reported Ti$_3$C$_2$ MXene (30.6%)[25] and antimonene quantum dots (AMQDs, 45.5 %).[65]

In addition to the extinction coefficient and PTCE, the photothermal stability is an important property for PTT. **Figure 5a** shows six photothermal cycles at concentrations of 25 ppm and 50 ppm. In one photothermal cycle, the sample was irradiated for 10 min, followed by a further 10 min period when the laser was turned off. It was found that the temperature initially increased to almost saturation level and then decreased to room temperature by natural cooling (**Figure 5a**). For the 25 ppm concentration, the highest temperature throughout the six cycles was approximately consistent, illustrating that the Ti NSs did not appreciably deteriorate during the 2 h photothermal process. While for the 50 ppm sample, it was observed with interest that there was a rising trend in the highest temperature as irradiation cycle number increased, shown by the dotted line. This was due to the increasing concentration of the

Ti NSs following evaporation of the water dispersant, as a result of the strong photothermal heating. This was supported by the increased absorbance after 2h of irradiation (**Figure 5b**). However, it was unexpectedly found that there was a slight decrease in the absorbance after 2 h storage in water, indicating the degradability of Ti NSs in water. The degradability of Ti NSs was further demonstrated after 30 days. It was found that both the absorbance and photothermal temperature change decreased significantly after 30 days (**Figure 5a, b**). Quantitatively, absorbance measurements showed that 77% of the Ti NSs degraded, however the photothermal temperature only degraded by 47%. The degradation product was $TiO_2$, as established by the XPS characterization in **Figure 2.** $TiO_2$ may have a significant impact on the decrease in absorbance, but led to only a minor decrease of photothermal temperature because it can also contribute to the photothermal temperature rise.[66–68] The degradability of Ti NSs is another advantage of Ti NSs for biomedical application.

Consequently, the high extinction coefficient (20.8 $L.g^{-1}.cm^{-1}$), high photothermal conversion efficiency (73.4%), good photostability and degradability of Ti NSs make Ti NSs a unique and high efficiency photothermal agent. Their excellent biocompatibility and photothermal killing performance in the following studies further support these assertions.

## Toxicity assays

Ti metal is well known for its excellent biocompatibility, however the cytotoxicity of the newly fabricated Ti NSs had to be ascertained to support their further clinical application. The potential cytotoxicity of both the naked and PEG encapsulated Ti NSs was evaluated. PEG was chosen in light of its biocompatibility and has been approved for medical use by the FDA. Tumor microenvironments consist of various kinds of cells, including tumor infiltrating immune cells. To model the true tumor microenvironment, both normal and cancer cells were incubated with the Ti NSs, including SMMC-7721 (hepatocellular carcinoma), B16 (melanoma) and J774A.1 (macrophage). As shown in **Figure 6a,b**, both naked and PEGylated Ti NS-treated cells exhibited trace cytotoxicity in CCK-8 assays post Ti NSs treatment; even at high concentration (100 ppm) which contrasts with the concentration of 30 ppm required for the total cell killing effect in the following study as shown in **Figure 6**.

The *in vivo* toxicity of both Ti NSs and PEGylated Ti NSs was also investigated. Given the proof of concept objective of this study, we used subcutaneous injection of Ti NSs for the toxicity assays study. Three groups of mice were used, which were injected with naked Ti NSs, PEGylated Ti NSs or saline. **Figure 6c** shows that the body weights of the mice dosed with Ti NSs or PEGylated Ti NSs showed no difference to those injected with saline, which suggests that the Ti NSs did not intrinsically affect the overall condition of the animals. In addition, H&E staining of the major organs including heart, liver, spleen, lung and kidney, showed no damage resulting from injecting the Ti NSs analogues (**Figure 6d**).

Consequently, our results demonstrate that the fabricated Ti NSs were non-toxic both *in vitro* and *in vivo*, and were therefore biocompatible, supporting further biomedical development.

## *In vitro* photothermal experiments

Based on the high efficiency photothermal performance and biocompatibility of Ti NSs, high tumor cell killing efficiency of Ti NSs was anticipated. As shown in **Figure 7a, b**, both Ti NSs and PEGylated Ti NSs showed a clear photothermal killing effect as concentration increased. 75% of SMMC-7721 and B16 cells were killed at 20 ppm and almost all of the cells were killed at a low concentration of 30 ppm. J774A.1 cells were particularly susceptible, with only ~25% of the cells surviving after the photothermal killing process, even at 10 ppm. This could be because J774A.1 cells can take up larger amounts of Ti NSs than cancer cells due to phagocytosis (data not shown), resulting in higher photothermal efficiency.

Additionally, we investigated the laser irradiation time required to achieve 100% cell killing efficacy at 50 ppm. It was found that a short time of 2 min of laser irradiation killed all of the cells. We attribute the efficient cell killing, to the sharp increase in temperature, passing the fatal point, induced by NIR irradiation (**Figure 7c**). Graphic illustrations of the photothermal killing effect for different concentrations of Ti NSs are presented in **Figure 7d**. We demonstrated the high photothermal killing effect of Ti NSs, and their excellent biocompatibility, making them promising for further *in vivo* development for cancer therapy.

## 3. Conclusions

In this work, we found that non-layered metallic Ti can be fabricated into 2D nanosheets (NSs) using liquid-phase exfoliation (LPE), with an average thickness of ~3 nm, and lateral size of less than 50 nm. The ultrathin Ti NSs exhibited a high extinction coefficient of 20.8 $L.g^{-1}.cm^{-1}$. They also showed a good photothermal conversion efficiency (PTCE) of 73.4% under NIR irradiation (808 nm laser), owing to the localized surface plasmon resonances (LSPR) resulting from the metal nature of the Ti NSs. This PTCE is significantly higher than those of classic photothermal agents such as Au (21%) and emerging 2D NMs, such as $MoS_2$ (24.4%), black phosphorus (28.4%) and $Ti_3C_2$ MXene (30.6%). The Ti NSs also degraded in a relatively short time (77% degradation over 30 days in water at ~30 °C). Thanks to the oxidization of Ti NSs, the Ti NSs presented the hydrophilic feature. In addition to their high photothermal performance, the as-prepared Ti NSs showed trace amounts of toxicity both *in vitro* and *in vivo*, and could also be encapsulated using conventional coating materials such as PEG, to improve their water dispersibility. As expected, the Ti NSs exhibited excellent therapeutic effects in a photothermal tumor therapy study, in hepatocellular carcinoma models. Consequently, the widely used biomedical material of Ti was demonstrated as a promising photothermal agent with excellent biocompatibility and high photothermal efficacy. Given the strengths of LSPR for elemental metals, our study would arouse a wide interest for PTT using other elemental transition metal.

## 4. Experimental section

### Materials

Commercially available Ti powder was purchased from Macklin Company. DSPE-PEG, 5000 Da was purchased from Nanocs Inc. (New York, USA). All cell lines were obtained from American Type Culture Collection (ATCC). Acridine orange (AO) and propidium iodide (PI) assay kits were obtained from Logos Biosystems. Ultrapure water (18.25 MΩ.cm, 25 °C) was used to prepare water-based dispersions. All other reagents used in this work were analytical reagent grade.

### Synthesis of Ti NSs

The ultrathin Ti NSs were prepared from non-layered bulk Ti using liquid-phase exfoliation. Typical exfoliation was mainly divided into two steps: probe sonication and bath sonication. 500 mg of Ti powder was dispersed in 100 ml of IPA. The suspension was then subjected to probe sonication for 10 h at a power of 240 W. To avoid over heating during the sonication process, the sonication was set to an on/off cycle of 2/4 seconds and the Ti dispersion was kept in ice water. Subsequently, the Ti dispersion underwent bath sonication at a power of 360 W for 10 h. The water bath temperature was controlled at 10 °C.

After sonication, the resulting dispersions were centrifuged at 2000 ×g for 30 min to remove the un-exfoliated component. The supernatant containing the Ti NSs was decanted gently and then centrifuged for a further 30 min at 12000 ×g. The precipitate was dried in a vacuum drying oven. To avoid oxidation, the Ti NSs were packaged in tinfoil and stored at 4 °C in the refrigerator for further characterization or use in bio-experiments.

The PEGylated Ti NSs were further prepared. 1 mg of DSPE-PEG was dispersed in 1 ml water. 5 ml Ti NSs dispersion in water with concentration of 100 ppm was involved sonication for 30 min and then mixed with PEG solution. The mixture underwent bath sonication for several min and stir for 3 h. Then, in order to remove the excess PEG molecules, the resulting mixture was ultrafiltered in Amicon tubes (MWCO 100kDa; Millipore) at 1000 ×g until all water was filtered out, and was washed 2 times using the same method. The pure PEGylated Ti NSs were re-suspended in ultrapure water or culture media for further use.

## Characterization

To confirm the three-dimensional morphology of Ti NSs, both atomic force microscopy (AFM, Bruker, Dimension Fastscan) and transmission electron microscopy (TEM, JEM1230) were used. AFM samples were prepared by dispersing on a silicon substrate using the drop-casting method and AFM images were scanned at 512 pixels per line. High-resolution TEM images and selected-area electron diffraction (SAED) were obtained using a Tecnai G2 F30 with an acceleration voltage of 300 kV. X-ray diffraction (XRD) patterns were acquired using a Philips X'Pert Pro Super diffractometer. X-ray photoelectron spectra (XPS) were acquired using a VG Escalab MK II spectrometer. UV-Vis-NIR absorption spectra were measured in the range 400–1000 nm using a Cary 60 spectrometer from Agilent. The Fourier transform infrared (FTIR) spectra were measured to verify the PEG coating of Ti NSs. An 808 nm fiber-coupled continuous semiconductor diode laser, LSR808H from Lasever Inc., was used as the laser source for the photothermal experiments. For the photothermal temperature measurements, an infrared thermal imaging camera (FLIR E-60) and thermocouple were used.

## Cell culture assays

Mouse melanoma B16 cells and mouse macrophage J774A.1 cells were cultured in Dulbecco's Modified Eagle Medium (DMEM) with high glucose (Hyclone). Human hepatocellular carcinoma SMMC-7721 cells were maintained in a 1:1 mixture of DMEM and Ham's F-12 medium (Hyclone). All of the culture media were supplemented with 10% fetal bovine serum (Gibco) and 1% Pen/Strep (Gibco). Cells were cultured in an incubator (Thermo Fisher Scientific) at 37 °C with 5% $CO_2$.

## *In vitro* experiments

SMMC-7721, B16 or J774A.1 cells were incubated in 96-well plates. After adherence (~12 h later), cells were treated with different culture media containing naked Ti NSs or PEGylated Ti NSs, or remained untreated (Mock). For *in vitro* cytotoxicity assays, cells were directly subjected to cell-counting kit (CCK8) assays (Beyotime Biotechnology) 24 hours post Ti NSs incubation. For *in vitro* photothermal experiments, cells were incubated with Ti NSs for 4 h and were then irradiated with an 808 nm laser, 1 $W.cm^{-2}$ for 10 min. After a further 24 h, cells were subjected to CCK8 assays according to the manufacturer instructions. The relative cell viability was normalized to the Mock samples (concentration of Ti NSs = 0 ppm) of each cell line. For fluorescent imaging of photothermal effects, cells were treated with the indicated concentrations of PEGylated Ti NSs for 4 h, and were subsequently irradiated as described above. 24 h post irradiation, cells were subjected to calcein AM/PI staining (Sigma), live cells (green, stained by calcein AM) and dead cells (red, stained by PI) were revealed.

## Mouse experiments

All of the animal studies were performed in compliance with the guidelines approved by the Animal Welfare and Research Ethics Committee at Shenzhen University (ID: 2017003). The mice used in this work were purchased from Guangdong Medical

Laboratory Animal Center (Guangzhou, China). Mice were euthanized before the ACUSU maximum allowable tumor burden of 2 cm$^3$.

## *In vivo* toxicity experiments

For all *in vivo* studies, Ti NSs or PEGylated Ti NSs were dispersed in saline. 6-week-old female Balb/c nude mice were randomly divided into 3 groups (n=5 each group). Mice were subcutaneously injected with saline, Ti NSs or PEGylated Ti NSs at day 1. The dose of Ti NSs or PEGylated Ti NSs was adjusted to 0.25 mg.kg$^{-1}$ with an injection volume of 100 μl and concentration of 50 ppm. To monitor the *in vivo* toxicity, body weight was measured every 2 days until day 15. The mice were then euthanized and the major organs, heart, liver, spleen, lung and kidney were collected for H&E staining as previously reported.[69]

## *In vivo* photothermal tumor therapy

To study the PTT efficiency of PEGylated Ti NSs in treating tumor tissue, human hepatocellular carcinoma models were established by subcutaneous injection of 5×10$^6$ SMMC-7721 cells in the left flank of Balb/c nude mice (6-week-old, female). Approximately 10 days post injection, the tumor volumes reached 100–200 mm$^3$. The tumor bearing mice were randomly divided into 4 groups (n=5 each group) for different treatments by intratumoral injection: *group 1*, saline; *group 2*, PEGylated Ti NSs; *group3*, saline with NIR irradiation; *group 4*, PEGylated Ti NSs with NIR irradiation. To achieve an injection dose of 0.25 mg.kg$^{-1}$ of PEGylated Ti NSs, the injection volume was 100 μl of 50 ppm in saline.

6 h post injection, the mice were anaesthetized and the tumors were irradiated with the NIR laser (808 nm, 1 W.cm$^{-2}$, 5 min). The distance between the laser point and tumor was 6 cm, and the temperature of the tumor was monitored using an infrared thermal imager. Following laser irradiation, the body weights and tumor volumes were recorded every 2 days. Tumor dimensions were measured using a caliper and the volumes were calculated using the formula (volume=length×width$^2$/2, V=l×w$^2$/2). At day 15, all of the mice were euthanized and the major organs, heart, liver, spleen, lung and kidney were isolated for H&E staining to assess the possible damage caused by the PTT.

## Statistical analysis

All of the data were analyzed using Graphpad Prism software and are presented as means ± SD. For *in vitro* studies, cell viability was normalized to the mean of Mock samples for each cell line, which was set to be 100%. Analysis of significance was performed by student's *t*-tests. For animal experiments, the significance analysis of tumor volumes between the four groups was performed using multiple *t*-tests. *p*<0.05 was considered statistically significant, \**p*<0.05, \*\**p*<0.01, \*\*\**p*<0.001.

## Supporting Information

Supporting Information is available from the Wiley Online Library or from the author.

## Acknowledgements

This research is partially supported by the National Natural Science Fund (Grant Nos. 61435010 and 61575089), Science and Technology Innovation Commission of Shenzhen (KQTD2015032416270385 and JCYJ20150625103619275), and China Postdoctoral Science Foundation (Grant No. 2017M612730). We thank Sarah Dodds, PhD, from Liwen Bianji, Edanz Editing China (www.liwenbianji.cn/ac), for editing the English text of a draft of this manuscript.

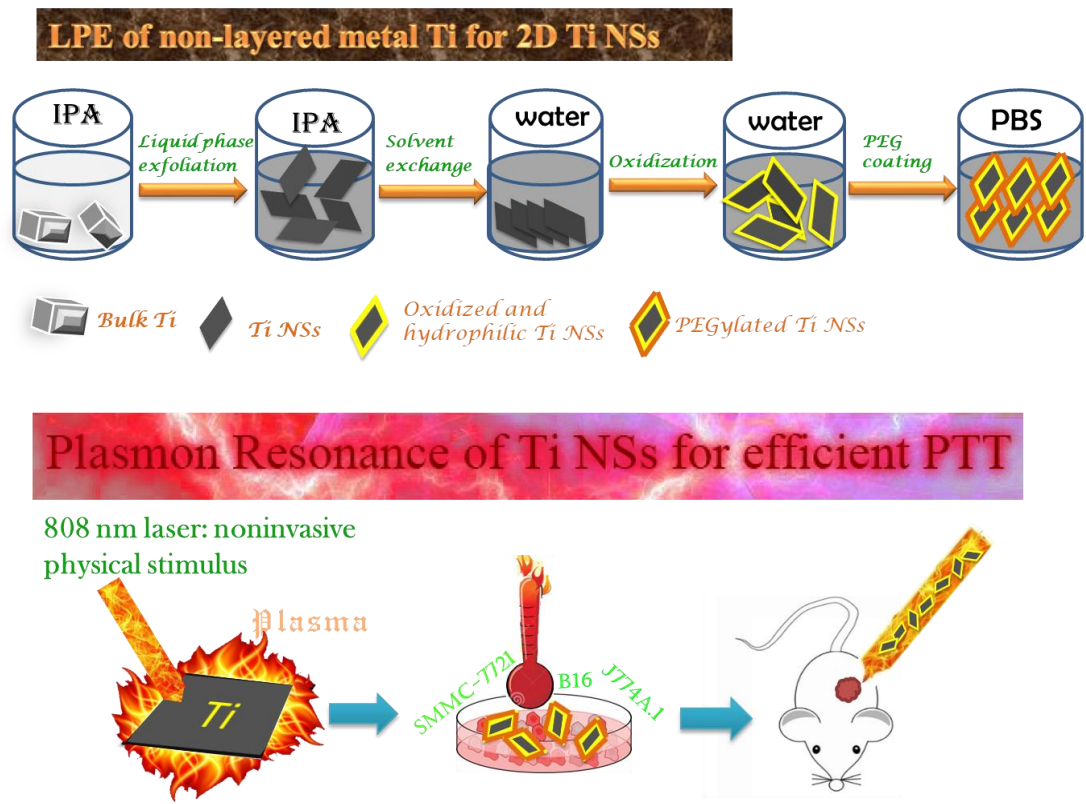

Figure 1. Schematic representation of the liquid-phase exfoliation (LPE) of Ti nanosheets (NSs) and Ti NSs based plasmonic photothermal therapy (Ti-PPTT) for *in vitro* and *in vivo* experiments.

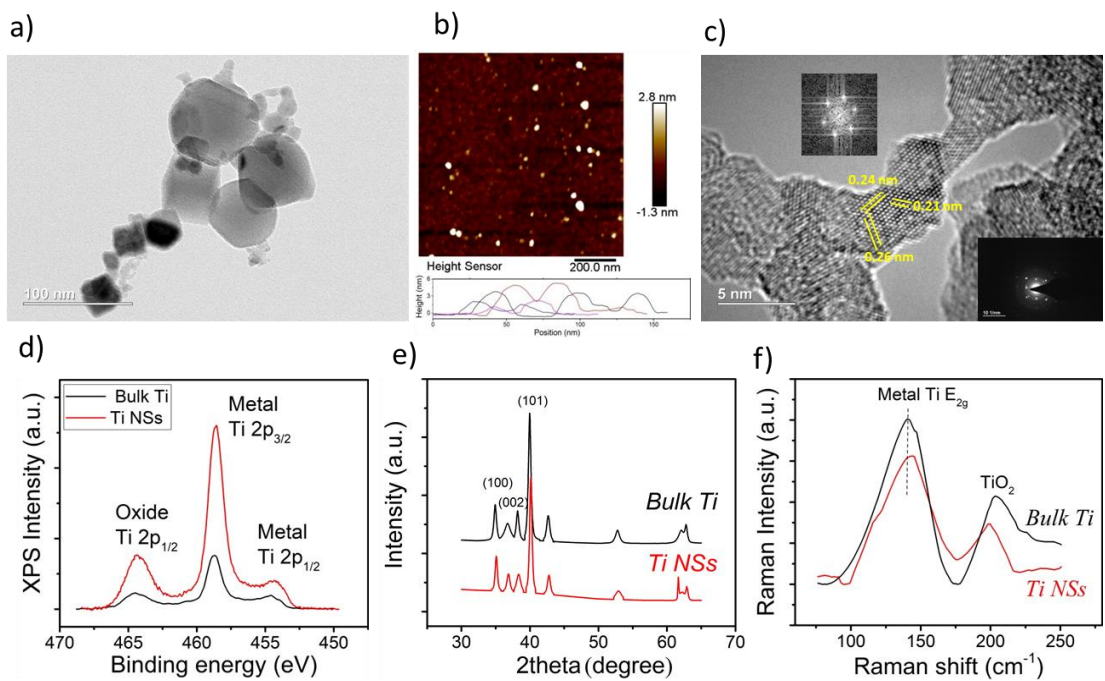

Figure 2. Typical characterization of exfoliated Ti NSs, including transmission electron microscopy (TEM), high-resolution transmission electron microscopy (HRTEM), atomic force microscopy

(AFM), X-ray photoelectron spectroscopy (XPS), X-ray diffraction (XRD) and Raman spectroscopy. a) TEM and b) AFM image of Ti NSs and Ti QDs. c) Crystal lattice, selected-area electron diffraction (SAED) and corresponding Fast Fourier Transformation (FFT) of Ti NSs. d), e) and f) XPS, XRD and Raman spectra of bulk and exfoliated Ti NSs.

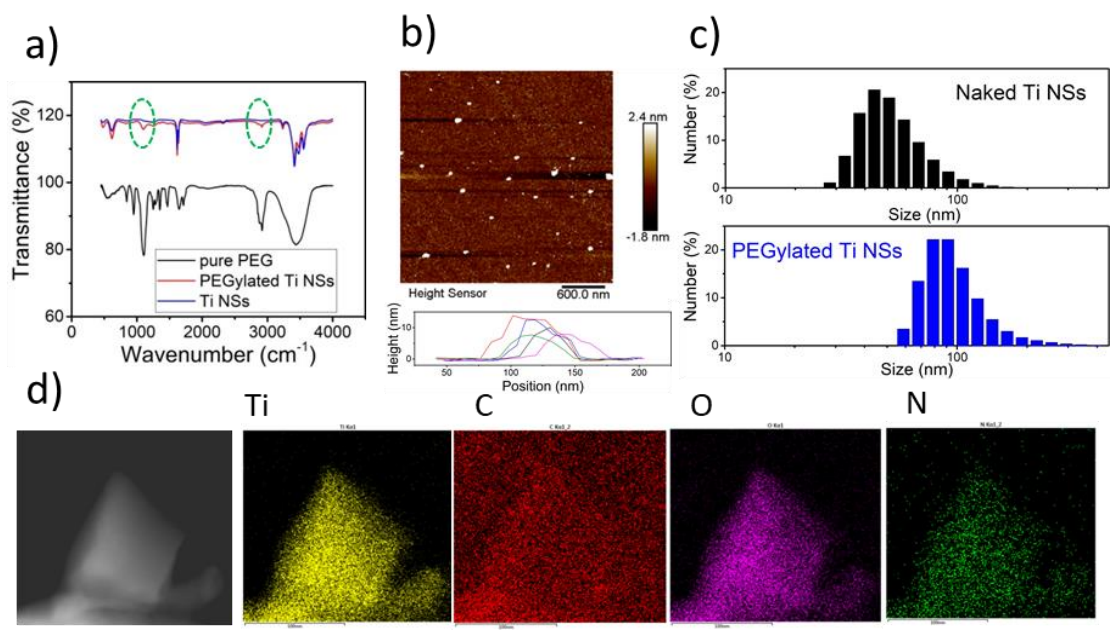

**Figure 3. Characterization of PEGylated Ti NSs.** a) FTIR spectra of pure PEG, PEGylated Ti NSs and naked Ti NSs. b) AFM image of PEGylated Ti NSs. c) Dynamic Light Scattering (DLS) size distribution of naked Ti NSs and PEGylated Ti NSs. d) STEM EDS mapping of PEGylated Ti NSs.

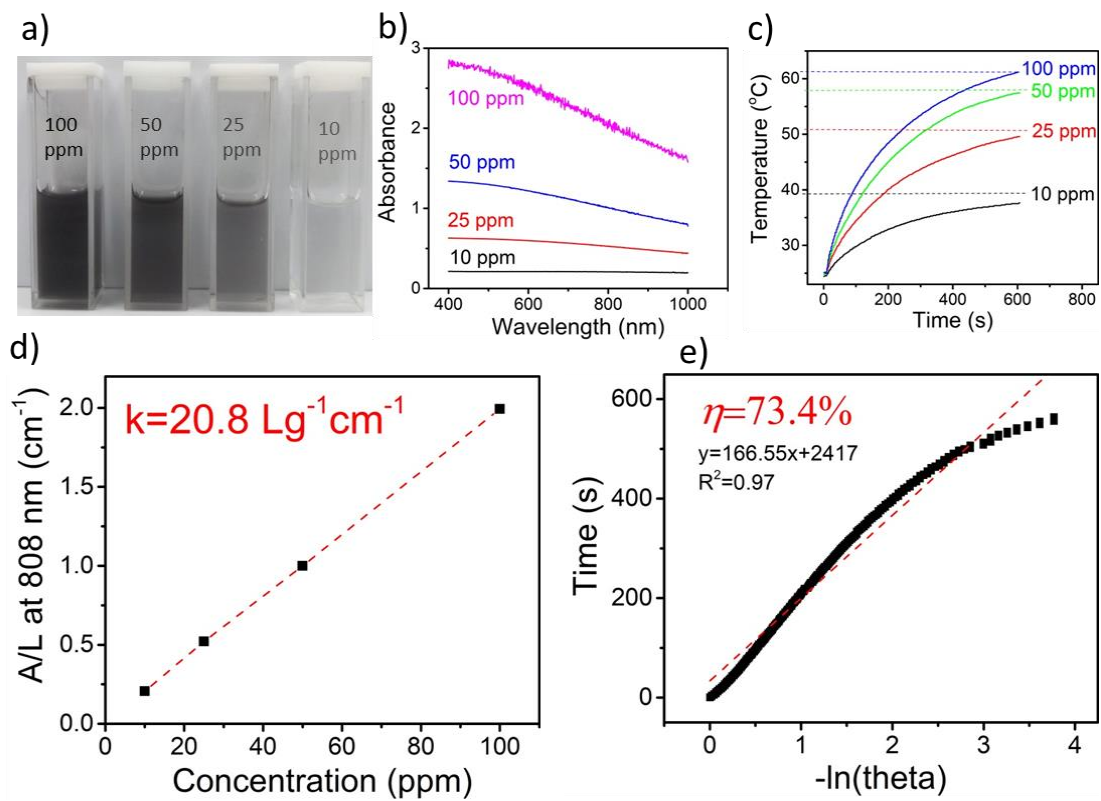

**Figure 4. Basic characterization of photothermal effect of Ti NSs.** a) Photograph of Ti NSs dispersed in water at different concentrations. b) Absorption spectra of Ti NSs. c) Photothermal temperature increase of water dispersed Ti NSs. d) The normalized absorbance intensity divided by the length of light path (A/L) at 808 nm. e), The linear fitting relationship between -ln θ and cooling time in one photothermal cycle, to determine the photothermal conversion efficiency (PTCE).

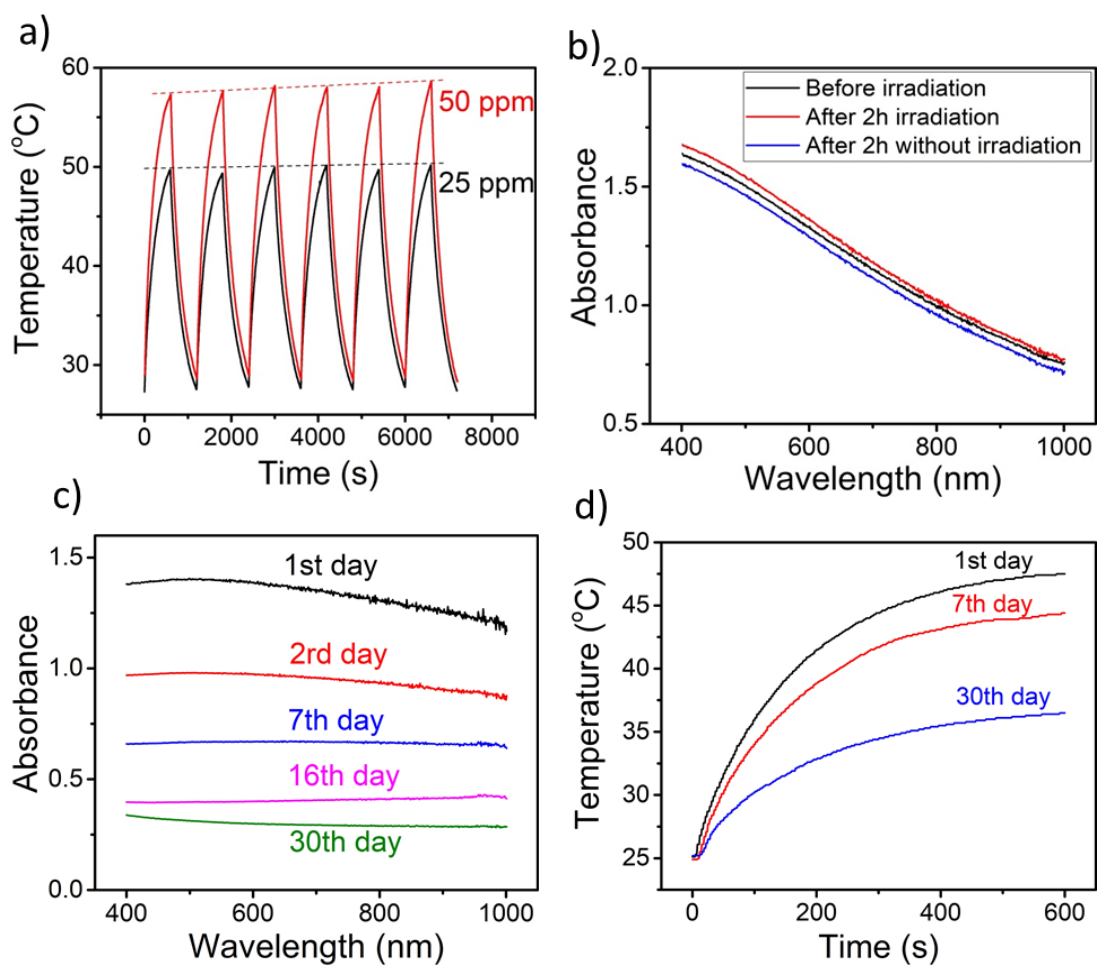

**Figure 5. Photostability and temporal stability of Ti NSs.** a) Photothermal stability for six cycles. The dotted lines show the relatively stable photothermal performance of Ti NSs. b) Laser stability and temporal stability of Ti NSs over 2 hours shown by absorbance. c) and d) Temporal stability of absorbance and photothermal temperature over 30 days.

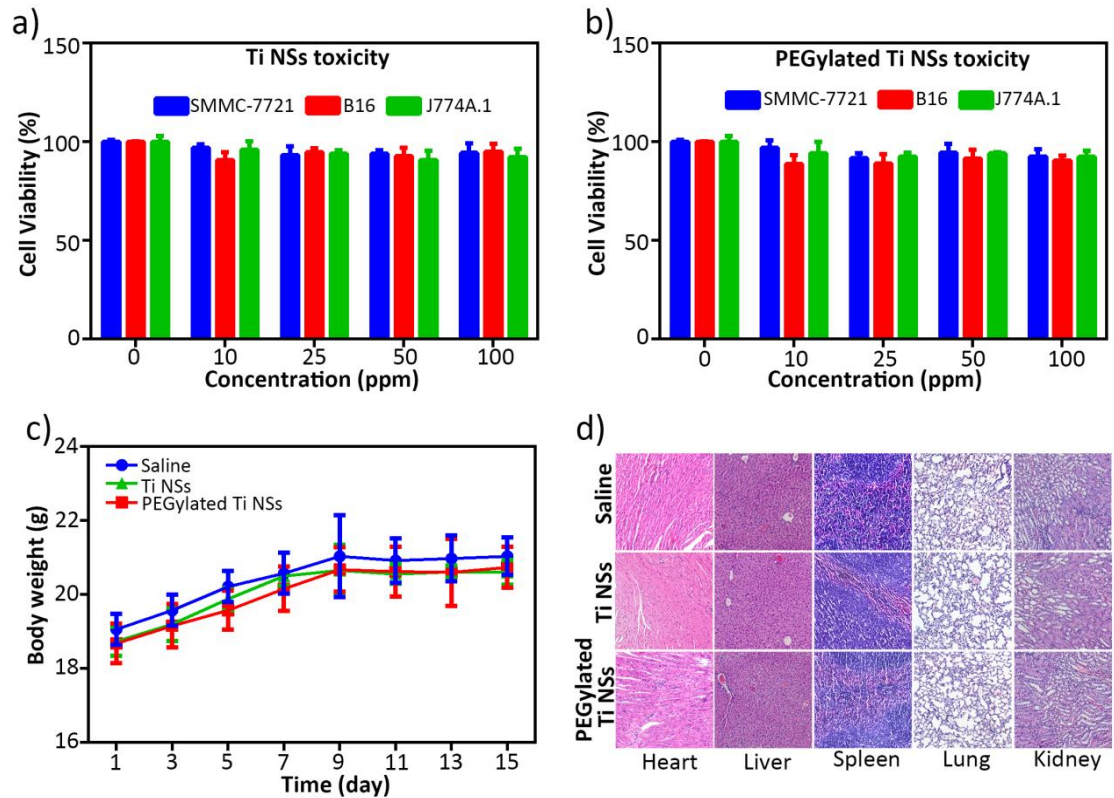

**Figure 6. Toxicity assays.** *In vitro* cytotoxicity of a) naked Ti NSs and b) PEGylated Ti NSs was assessed using SMMC-7721, B16 and J774A.1 cells. Ti NSs dispersed in the corresponding culture media with concentrations of 0, 10, 25, 50 and 100 ppm were incubated with the cells. c-d) *In vivo* toxicity. c) The body weight of mice measured at the indicated times and d) organ conditions assessed by H&E staining at day 15 post subcutaneous injection of saline, Ti NSs or PEGylated Ti NSs.

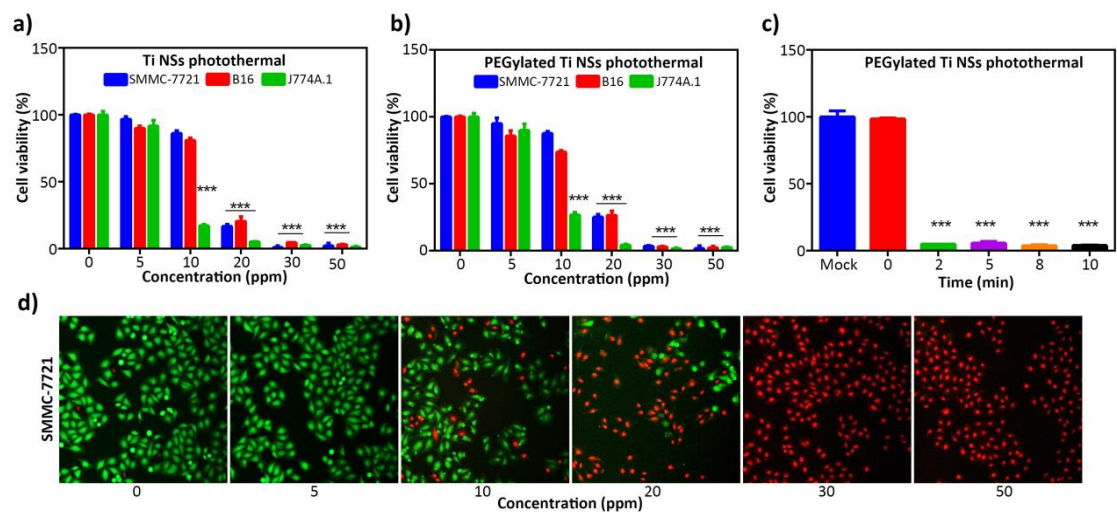

**Figure 7. *In vitro* photothermal experiments.** The photothermal cell killing effect of a) naked Ti

NSs and, b) PEGylated Ti NSs on SMMC-7721, B16 and J774A.1 cells for different concentrations under the same NIR irradiation conditions. Cells were incubated with Ti NSs or PEGylated Ti NSs at concentrations of 0, 5, 10, 20, 30 and 50 ppm for 4 h, followed by NIR laser irradiation (1 W.cm$^{-2}$, 10 min). c) Photothermal cell killing effect of PEGylated Ti NSs (50 ppm) for different irradiation times. SMMC-7721 cells were incubated with PEGylated Ti NSs for 4 h, followed by NIR laser irradiation (1 W.cm$^{-2}$) for different times (0, 2, 5, 8, 10 min). The Mock samples were untreated. d) Fluorescent image of photothermal effect on SMMC-7721 cells post irradiation. Cells were incubated with the indicated concentrations of PEGylated Ti NSs, followed by NIR irradiation (808 nm, 1 W.cm$^{-2}$, 10 min), and then stained with calcein AM/PI. n=3 biological replicates, *** $p<0.001$, *t*-tests.

**Titanium (Ti), a widely used biomedical material, is first demonstrated for photothermal cancer therapy (PTT) and with high efficiency and excellent biocompatibility.** Using liquid-phase exfoliation (LPE), the non-layered metal Ti can be fabricated into 2-dimentional (2D) Ti nanosheets (NSs). The 2D Ti NSs exhibited high extinction coefficient of 20.8 Lg$^{-1}$cm$^{-1}$ and high photothermal conversion coefficiency (PTCE) of 73.4%, owing to localized surface plasmon resonances (LSPR). Our study could arouse a wide interest in the potential efficient PTT for other elemental transition metals owing to their common LSPR.

*Zhongjian Xie[a,1], Shiyou Chen[a,1], Quan Liu[c,d], Zhitao Lin[a], Jinlai Zhao[d], Taojian Fan[a], Dou Wang[c,d], Liping Liu[b,*], Shiyun Bao[b], Dianyuan Fan[a], and Han Zhang[a,*]*

# Biocompatible Two-dimensional Titanium Nanosheets for Efficient Plasmonic Photothermal Cancer Therapy

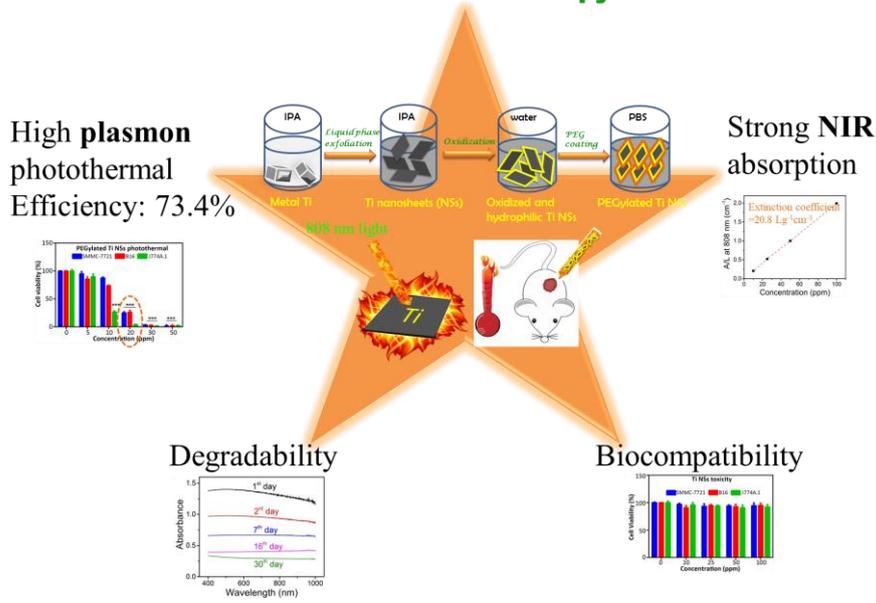